\input{aipcheck}

\documentclass[
    ,final            
  ]
  {aipproc}

\layoutstyle{6x9}

\newcommand{\beq}{\begin{equation}}
\newcommand{\eeq}{\end{equation}}
\newcommand{\bea}{\begin{eqnarray}}
\newcommand{\eea}{\end{eqnarray}}
\newcommand{\ben}{\begin{eqnarray*}}
\newcommand{\een}{\end{eqnarray*}}

\newcommand{\lambdaqcd}{\Lambda_{\rm QCD}}

\def\D0{D\O}

\begin{document}

\title{Heavy-quark recombination in $Z^0$ decay}

\author{Yu Jia}{
  address={Department of Physics and Astronomy, Michigan State University \\
East Lansing, MI 48824}
}

\begin{abstract}
We briefly review the recent advances of heavy-quark recombination mechanism. 
This mechanism predicts a class of power-suppressed 3-jet events in $Z^0$ decay,
such as $b\overline{b}q$ and  $b\overline{b}\overline{q}$.
Furthermore, heavy quark fragmentation function also receives a contribution
from this mechanism. Some light can be shed on the scaling of
the maximum of the
fragmentation function for S-wave heavy hadrons.
We finally comment on a new variant of this mechanism 
which has important impact on the precision electroweak physics.
\end{abstract}

\maketitle



Heavy flavor production serves as an excellent testing ground for perturbative
QCD~\cite{Frixione:1997ma}. So far, the heavy quark cross sections
in all different processes have been computed 
to at least the next-to-leading order. 
However, in order to compare with experimental data, 
a sound understanding of  how a heavy quark turns into a heavy hadron 
is crucial.  
The standard strategy is to implement the heavy quark fragmentation
as the sole hadronization  mechanism.

Inspired by the Non-relativistic QCD factorization for 
the formation of heavy quarkonium~\cite{Bodwin:1994jh},
a new hadronization mechanism,  dubbed {\it heavy-quark recombination} 
(HQR) was recently developed~\cite{Braaten:2001bf}.
It was initially motivated as a ``higher twist" mechanism, 
to supplement the usual fragmentation.
The central picture of this mechanism is quite simple: 
after a hard scattering, 
a heavy quark may capture a nearby light {\it parton} which emerges from
the hard scattering and happens to carry soft momentum in
the heavy quark rest frame. Subsequently they can materialize into a heavy hadron, 
plus additional soft hadrons.
A typical HQR process in hadron collision at $O(\alpha_s^3)$
is $\overline{q}\,g\rightarrow \overline{B}+\overline{b}+X$, where
$\overline{B}$ is produced from the $b\overline{q}$ recombination.
Similarly, $cq$ recombination has later been introduced 
to account for $\Lambda_c$ production~\cite{Braaten:2003vy}.

HQR is drastically distinct from the other ``higher-twist"
mechanisms -- conventional recombination model~\cite{Das:1977cp}, 
intrinsic charm model~\cite{Vogt:1995fs}, 
and so on. In all these cases,
the beam remnants participate in the dynamics of forming a heavy hadron.
In contrast, the beam remnants play no role in HQR processes, which
leads to great simplifications. 
In fact, HQR respects a simple factorization formula.
Namely, inclusive production of heavy hadron in HQR can be expressed as
a product of hard-scattering parton cross section, which is calculable
in perturbative QCD, and  a nonperturbative parameter 
({\it recombination factor}),
which characterizes the probability for the heavy quark and the light parton to 
evolve into a state containing the heavy hadron~\cite{Braaten:2001bf}. 

One important achievement of HQR 
is that it can explain 
the charm meson and baryon production asymmetries
observed in a number of fixed-target experiments,
in a simpler, more coherent
and controlled fashion than those 
aforementioned models~\cite{Braaten:2001uu,Braaten:2003vy}.
The charm asymmetry is simply attributed to the asymmetry between the 
densities of light quark and anti-quark in the beam and target hadron.

Although this success constitutes a strong evidence for HQR,
the complicated hadronic environment in fixed-target experiments
prevents us from excluding other hadronization models.
Most probably, the asymmetries arise from the interplay 
of several different mechanisms, one of which is HQR. 
A curious question  thereby is,
is there a cleaner playground where HQR can be unambiguously singled out?

The answer is {\it yes}, because the physical idea of HQR is quite general, 
so its applications are not only confined in the hadroproduction
of heavy hadron.
In fact, heavy flavor production in $e^+e^-$ annihilation 
is an ideal place to test HQR~\cite{Jia:2003ct,Jia:2003jk}.
In particular, we will be interested in $B$ production on the $Z^0$-pole,
thanks to the huge statistics of $Z^0$ samples.
Clearly, those ``higher-twist" mechanisms which rely on 
the beam remnants in hadron collision,
are simply absent here.

\begin{figure}\label{zbb}
  \includegraphics[height=.14\textheight]{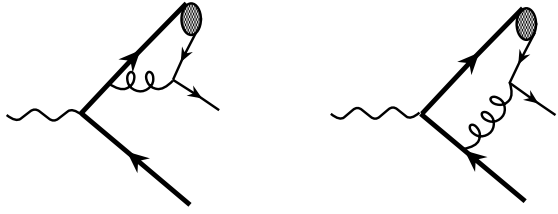}
  \caption{
Diagrams for the $b\overline{q}$ recombination process
$ Z^0 \rightarrow b\overline{q}(n) + \overline{b}+q$. 
The shaded blob represents the hadronization of $b \overline q(n)$
into $\overline B$ meson plus anything else.
}
\end{figure}

Let us consider $B$ production at $O(\alpha_s^2)$ 
through $Z^0 \rightarrow b \overline{b} q\overline q$.
If each quark independently fragments, then it represents a regular 4-jet event.
Nonetheless, in a small corner of phase space where $\overline q$ is soft in 
the $b$ rest frame,
they can form a composite $b\overline{q}$ state with definite color and angular momentum. 
Subsequently this state hadronizes into a $\overline B$ meson plus soft hadrons.
We thereby end up with a  jet containing $\overline B$  from the recombination, 
the recoiling $\overline b$ jet and a light quark jet~\cite{Jia:2003ct}. 
The corresponding Feynman diagrams are shown in Fig.~\ref{zbb}. 
The inclusive $\overline B$ production rate from HQR can be written
\bea
\label{factorization}
d \Gamma [\overline{B}] = \sum_n d{\hat\Gamma}[Z^0\rightarrow b\overline{q}(n) + \overline{b}+q]\,
\rho[b\overline{q}(n) \rightarrow \overline{B}] \,. 
\eea 
%
where $d{\hat \Gamma}_n$ are the perturbatively calculable parton cross sections, 
and $\rho_n$ are the recombination factors, and $n$ denotes the color and angular momentum 
quantum numbers of $b\overline{q}$.
These $\rho_n$ parameters have recently been defined 
in terms of nonperturbative QCD matrix elements~\cite{Chang:2003ag}.
An important property of these parameters is that they scale as $\lambdaqcd/m_b$.
While $\overline B$ can be produced in four different 
recombination channels,
the color-singlet, spin-matching channel is expected to dominate.
Adopting the fitted value of $\rho_1^c$ from Ref.~\cite{Braaten:2001uu,Braaten:2003vy}, 
and using its scaling property, we can obtain $\rho_1^b=0.1$.

A striking signal of these novel 3-jets is that the third jet 
is initiated by a {\it light quark}, instead of by a {\it gluon}. 
However, distinguishing quark and gluon jets experimentally
requires a large statistics. 
OPAL collaboration has selected 3,000 symmetric 3-jet events at the $Z$ resonance,
in which the most energetic jet is tagged to contain $b$, and 
the angle between this $b$-jet and each of
the two low energy jets is roughly $150^\circ$~\cite{Alexander:1995bk}.
These samples were assumed to all be the $b \overline{b} g$ events.
However now we know there must be a small fraction of them
are actually  made of HQR 3-jets.
Simple dimensional argument suggests these 3-jets are suppressed 
by a factor of $\alpha_s(M_Z)\lambdaqcd m_b/M_Z^2\sim 10^{-5}$ relative to 
$b \overline{b}g$. However, a more quantitative study indicates
that the ratio of the yield for HQR 3-jet events
to that for the $b \overline{b} g $  in such a topology is
roughly $0.012$.
So there are about 36 new events out of 3,000 OPAL samples,
seemingly not statistically important.
We hope that prospective Giga-Z experiments with a much larger number of
$Z^0$ samples will confirm the existences of these 3-jet events definitely.
If true, it should be viewed as a decisive triumph of the 
HQR mechanism.

Though the HQR cross section is highly suppressed for 3-jet,
its magnitude becomes much larger when $\overline{B}$ and $q$ lie 
in the fragmentation region, i.e., with a small invariant mass,
because the virtuality of the internal gluon that splits into $q\overline{q}$ 
(see Fig.~\ref{zbb}) becomes much smaller in this region.
This motivates us to examine if this $b\overline{q}$ recombination
process also contributes to the $b$ fragmentation function.

Fragmentation functions are nonperturbative objects
and usually defy a tackle from perturbation theory.
This is true for $q$, $g$ to fragment into $\pi$, $K$.
However, the fact that $b$ is heavy ($m_b \gg \lambdaqcd$)
may allow us to proceed further.
Armed with knowing how a heavy quark hadronizes in
the recombination picture encoded in Eq.~(\ref{factorization}),
we can readily derive the HQR contribution to
$b$ fragmentation function
by integrating the inclusive $\overline{B}$ differential cross sections 
over some appropriate kinematic variables.
For example, the HQR contribution to $b$ 
fragmentation into $\overline{B}^*$ turns out to be
\bea
D_{b \rightarrow \overline{B}^*}^{\rm HQR}(z) &=&
{32 \, \rho_1^b \, \alpha_s^2(m_b)\over 81}
\,{z\,(2-2\,z+3\,z^2) \over (1-z)^2}\,,
\label{frag}
\eea
where $z$ is the energy fraction carried by $\overline{B}^*$
relative to $b$.
This HQR fragmentation function is not away from zero
until $z$ becomes large, and finally diverges quadratically
as $z\to 1$.
This divergence is a symptom that perturbative calculation
in the endpoint region becomes invalid. Yet, one can show
that Eq.~(\ref{frag}) is still valid as long as $1-z \gg  \lambdaqcd$.
The $z$ distribution in Eq.~(\ref{frag}) is much harder than the
widely-used Peterson fragmentation function~\cite{Peterson:1983ak}. 
This may suggest
that $b$ hadronizing via picking up a 
$\overline{q}$ from vacuum is still non-negligible,
even at relatively small $z$. 
However, some 
model-independent extraction of the nonperturbative part of
$b$ fragmentation function  shows also a harder spectrum
than Peterson parameterization~\cite{Ben-Haim:2003yu}.

Insight may be gained if we assume that $z \sim 1- \lambdaqcd$ is 
where the peak of fragmentation
function is located. While a perturbative QCD 
treatment from HQR is ceasing to work when close to
the endpoint  region, 
$D^{\rm HQR}_{b\rightarrow \overline{B}^*}(1-\lambdaqcd/m_b)$ 
may still betray the correct order of magnitude of the maximum of the ``true" 
fragmentation function. 
If this is true,
then the fragmentation function of $b$ to $B^{*-}$ 
is expected to peak around
$z=0.93$, with a height roughly 
${32 \over 27} \,\rho_1^b \,\alpha_s^2(m_b) (m_b / \lambdaqcd)^2 \approx 1.5$.
If we approximate the ``true" $B^{*-}$ fragmentation function
by Peterson function $D(z;\epsilon_b)$  
with $\epsilon_b=0.006$, and take the fragmentation probability 
$f_{b\rightarrow B^{*-}}\approx 0.3$,
the ``true" peak is also around $z \approx 0.93$ with a height about 1.7, in good agreement with  
our naive estimate. 
Since $\rho_1^b \propto \lambdaqcd/m_b$,  
we  thereby propose 
the maxima of the fragmentation functions for S-wave heavy hadrons
scale as $\alpha_s^2(m)\,m /\lambdaqcd$. For charmed hadrons, 
this scaling law doesn't hold so well, but still 
conveys the correct order of magnitude.

A comprehensive understanding of $Z^0$ decay to heavy flavor  
is important to
precision electroweak physics~\cite{Hagiwara:fs}.
If we were able to extract the finite power correction from
the linearly divergent total HQR cross section,
it would represent an $O(\alpha_s^2 \lambdaqcd m_b/M_Z^2)\sim 10^{-6}$
correction to the partial width of $Z^0$ to $b\overline{b}$. 
The $Z^0$ width has been measured to per mille accuracy, 
thus the contribution associated with Fig.~\ref{zbb}
can be neglected.

\begin{figure}\label{zbgb}
  \includegraphics[height=.12\textheight]{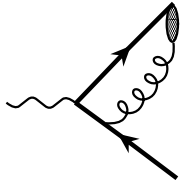}
  \caption{
The diagrams for the $bg$ recombination process
$ Z^0 \rightarrow bg(n) +\overline{b}$. }
\end{figure}

However, there is a new HQR process, as depicted in Fig.~\ref{zbgb},
occurring at order $\alpha_s$ only, with a genuine ``higher twist"
contribution of order $\lambdaqcd/m_b$~\cite{Jia:2003jk}. 
To accomplish this, $bg$ recombination needs to be invoked.
The net contribution of this new mechanism to the partial width of 
$Z^0$ to $b\overline{b}$  turns out to be $\Delta \Gamma [b\overline{b}]  =
32 \pi\alpha_s(M_Z) \xi^b_{3}/9 \,\Gamma_0[b\overline{b}]$,
where $\xi_3^b$ is an unknown color-triplet recombination factor.
Both $\xi_3^b$ and $\xi_3^c$ may be fitted from the global electroweak analysis, 
and consequently
the Standard Model predictions of various electroweak observables will 
be updated.

All the three HQR mechanisms, $b\overline{q}$, $bq$ and $bg$
recombination have now been fulfilled. 

\begin{theacknowledgments}
I thank E.~Braaten and T.~Mehen for 
the collaboration on laying down the foundation of 
heavy-quark recombination mechanism.
This work is supported in part by the 
National Science Foundation under Grant No. PHY-0100677.
\end{theacknowledgments}

\end{document}